\documentclass[preprint,aps,prc,showpacs,nofootinbib,secnumarabic,floatfix]{revtex4-1}

\usepackage{graphicx}
\usepackage{bm}
\usepackage{color}
\usepackage{gensymb}
\usepackage{amsmath}
\usepackage{slashed}
\usepackage{appendix}
\usepackage{hyperref}

\def\bfg #1{{\mbox{\boldmath $#1$}}}
\begin{document}











\title{
The effect of $T$-invariance violation in scattering of polarized $^3$He nuclei on tensor-polarized deuterons
}



\author{
  Yu.\,N.\, Uzikov$^{+\times\circ}$
\footnote{e-mail: uzikov@jinr.ru},
M.\,N.\,Platonova$^{*+}$\footnote{e-mail:platonova@nucl-th.sinp.msu.ru}}





\affiliation{
~\\$^+$
V.P. Dzhelepov Laboratory of Nuclear Problems, Joint Institute for Nuclear Research, Dubna, 141980 Russia
\\~\\
$^*$ D.V. Skobeltsyn Institute of Nuclear Physics, M.V. Lomonosov Moscow State University, Moscow, 119991 Russia
\\~\\
$^\times$ M.V. Lomonosov Moscow State University, Faculty of Physics, Moscow,
119991 Russia
\\~\\
$^{\circ}$ Dubna State University, 141980 Dubna, Russia}


\begin{abstract}
In the interaction of a transversely polarized nuclear beam
with a tensor-polarized deuteron target, a nonzero value of the component
of the total cross section of the process corresponding to this combination of polarizations
is an unambiguous
signal of $T$-invariance violation while $P$-parity is preserved. The method developed earlier
for calculating this component of the total
cross section for $pd$ scattering based on the Glauber theory has been generalized by us to the case
of $^3$He$d$ scattering, and its energy dependence
in the range of beam energies 0.1--1 GeV/nucleon has been calculated.
It is found that in $^3$He$d$ collisions, in contrast to $pd$ scattering, the contribution of only
one type of 
$T$-violating
nucleon-nucleon forces dominates, which
is essential for 
extraction of the unknown constant of this interaction from the corresponding data.
\end{abstract}

%

\maketitle

\textbf{1. Introduction.}

Discrete symmetries with respect to
space inversion ($P$),
time reversal ($T$) and charge conjugation ($C$)
play a key role in the theory of fundamental interactions
\cite{Vergeles:2022mqu}.
The violation of $C$- and
$CP$-symmetries is required to
explain the baryon asymmetry of the Universe
\cite{Sakharov:1967dj}.
Within the framework
of the Standard Model (SM) of fundamental interactions and the standard cosmological model, $CP$-violation
 observed in the physics of kaons, $B$- and $D$-mesons, is far from sufficient to explain this asymmetry -- it lacks 8--9
orders of  magnitude \cite{Riotto:1999yt}, \cite{Bernreuther:2002uj}.
It follows from this that there must be other sources of $CP$-violation in nature, beyond the
SM. All detected $CP$-violating effects (under the condition of $CPT$-symmetry, these effects are equivalent to the violation of $T$-invariance) 
simultaneously violate $P$-parity.
 The signal of the $T$-odd $P$-odd effects is
the permanent electric dipole moment (EDM) of the neutron, neutral atoms and charged particles
\cite{Engel:2013lsa}.
A lot of work has been devoted
to search for the neutron EDM,
as well as 
that of
charged particles -- protons, deuterons, 
and $^3$He nuclei
(see \cite{CPEDM:2019nwp} and references therein).

On the contrary, much less attention 
has been paid
to $T$-odd $P$-even, or Time-invariance Violating $P$-parity
Conserving (TVPC) and  flavor-preserving effects proposed in 1965 by Okun \cite{Okun:1965tu}, Prentki and Veltman \cite{Prentki:1965tt}, Lee and Wolfenstein \cite {Lee:1965hi} to explain the violation of
$CP$-symmetry.
In the framework of SM, these effects
are absent at the level of fundamental interactions and
can only appear as electroweak radiative corrections to $T$-odd $P$-odd interactions, while
their intensity is vanishingly small
 \cite{Gudkov:1991qg,Engel:1995vv}.
The reason why TVPC effects are of interest is that
the experimental constraints on them
are still quite weak, much weaker than on EDM, and the observation of the TVPC effect at
the currently achievable level of experimental accuracy
($\le 10^{-6}$)
will be a direct evidence
of physics beyond  the SM. Previously, it was assumed that the existing
experimental restrictions
on EDM simultaneously impose restrictions on TVPC interactions
\cite{Conti:1992xn}, while the expected magnitude of the effect becomes negligible.
However, it was later shown
\cite{Kurylov:2000ub},
that there is such a scenario
of EDM generation beyond  the SM, in which
there is no connection between experimental constraints on the
EDM and TVPC effects.

In the experiment, the search for $T$-invariance violation while $P$-parity 
is conserved is carried out
(see \cite{Barabanov:2004cr}
and references therein)
by verifying the principle of detailed balance in nuclear reactions,
measurements of $T$-odd angular correlations in
the beta-decay of nuclei and angular distributions of gamma-quanta, in violation of charge symmetry in the scattering
of polarized protons on neutrons and polarized neutrons on protons, in the transmission of polarized
neutrons through the aligned (tensor-polarized) nuclei \cite{Huffman:1996ix}.
Experimental limits on the value of
TVPC effects are consistently lowered.
So, the purpose of the experiment on $pd$ scattering at an energy of 135 MeV
\cite{Lenisa:2019cgb}
-- achieving the measurement accuracy 
of the
TVPC signal at the level of $10^{-6}$, which is an order
of magnitude better compared to the experiment on neutron transmission through a tensor-polarized target
of holmium nuclei, $^{165}$Ho
\cite{Huffman:1996ix}.

In the case of few-nucleon systems, it is possible to reliably calculate the absolute value and
energy dependence of the
TVPC signal with accuracy up to unknown constants of the $T$-violating
interactions that are included in the expression for the signal 
as factors. 
At collision energies characteristic for modern accelerators, the constants of
the TVPC interaction
with a high probability do not depend
on energy, so the 
shape of the energy dependence of the TVPC signal is determined by
the usual $T$-even $P$-even forces, and it is necessary to know this shape  in order
to choose an optimal energy region when
searching for the signal.
For $pd$ collisions at low energies 0.1--2 MeV, the energy
dependence of the TVPC signal was calculated based
on the solution of the Faddeev\cite{Song:2011jh} equations, and in the energy range 0.1--1 GeV --
based on the Glauber theory
\cite{Uzikov:2015aua,Uzikov:2016lsc}.
For the collisions of the $^3$He  nuclei with deuterons this problem
 is considered for the first
time in this paper on the basis of the corresponding modification
of the method  of Ref. \cite{Uzikov:2015aua,Uzikov:2016lsc}.

In the $S$-wave approximation for the wave function of the $^3$He nucleus, the polarization of this nucleus
is due to a neutron, so polarized beams of $^3$He nuclei are effectively beams of polarized neutrons and are of great interest for hadron spin physics
\cite{Accardi:2012qut}.
Recently, much attention has been paid to the creation of such beams
at the RHIC accelerator complex, the future electron-ion collider EIC \cite{Zelenski:2023kof}, as well
as the NICA complex \cite{Zelenski:2023dsp}.


\textbf {2. TVPC-signal.}
Consider the transmission of transversely polarized particles with spin $s=1/2$
through a target with tensor-polarized nuclei having spin $J=1$,
using as an example the experiment to test $T$-invariance in $pd$ scattering.
The polarization vector of the incoming proton is denoted
as ${\bf p}^p$, and the unit vector in the direction of the proton momentum is ${\bf m}$.
Let's choose a coordinate system so that
  OZ$ \uparrow\uparrow{\bf m}$, OY$\uparrow\uparrow {\bf p}^p$,
  OX$\uparrow\uparrow [{\bf p}^p\times {\bf m}]$.
  In general, the total cross section of the considered $pd$ interaction in
the presence of the effects of $T$- and $P$-invariance violation contains 9 terms
\cite{Nikolaev:2020wsj}. Provided that $P$-parity is preserved, this number 
is
reduced to five \cite{Uzikov:2015aua,Uzikov:2016lsc}, and if only
the transverse polarization of the proton beam ($p_y^p$)
is available, 
 is reduced to the following
four terms:
\begin{equation}
 \sigma_{tot}=\sigma_0+ \sigma_1 p_y^p  p_y^d+
 \sigma_3 P_{zz} +\sigma_{\small TVPC}p_y^pP_{xz},
 \label{sigmatot}
\end{equation}
 where $p_y^d$ is the transverse polarization of the deuteron, and $P_{zz}$ and $P_{xz}$ are
the components of the tensor  polarization  (alignment) of the deuteron.
The total cross section of the interaction
of the unpolarized proton and deuteron, $\sigma_0$, as well as the components of the
total cross section corresponding to the transversely polarized proton and deuteron,
$\sigma_1$, the unpolarized proton and
tensor-polarized ($P_{zz}$) deuteron, $\sigma_3$,
are due to the usual 
$T$-even $P$-even interactions.
%
The last term in the total cross section (\ref{sigmatot}),
$\sigma_{\small TVPC}$,
is caused by the interaction of the vector polarized protons ($p_y^p$) with tensor-polarized
deuterons ($P_{xz}$) and is a signal of violation of $T$-invariance while preserving $P$-parity.
This observable cannot be simulated by an interaction in the initial/final
states and is 
nonzero only if there is a TVPC-type interaction in the system.
This value is equivalent to the total cross section
of a fivefold $T$-odd
correlation introduced in \cite{Barabanov:1986sz}.

When the beam passes through the target, its intensity decreases due to absorption in the target, which is determined by the total cross section $\sigma_{tot}$.
To measure the cross section $\sigma_{\small TVPC}$ in an experiment with a fixed deuteron target,
it is necessary to measure the degree of attenuation of the beam as it passes
through the target for two opposite directions of proton polarization -- with spin up $T^{+}$ and down $T^{-}$ \cite{Huffman:1996ix,Lenisa:2019cgb}.
At zero value of the deuteron polarization vector, the asymmetry $(T^{+}-T^{-})/(T^{+}+T^{-})$ is proportional
to the cross section
$\sigma_{\small TVPC}$
\cite{Huffman:1996ix}.
If the transverse polarization of the deuteron $p_y^d$ is not zero, then, with this method of measuring the signal, the cross section $\sigma_1$ also contributes to the asymmetry, and this contribution is false.
Suppressing this contribution by suppressing
the absolute magnitude of the deuteron  polarization $p_y^d$ presents a serious problem.
So, in the case of $pd$ scattering at 135 MeV, the vector polarization of the deuteron $p_y^d$
should be suppressed to the level of $p_y^d < 2\cdot 10^{-6}$ \cite{Temerbayev:2015foa,Eversheim:2017zxl}
in order to ensure the planned accuracy of the experimental restriction on TVPC-signal
  $10^{-6}$ \cite{Lenisa:2019cgb}.
   The solution to this complex problem is provided by a new measurement method \cite{Nikolaev:2020wsj},
   based on the use of the deuteron beam polarization precessing in the plane of the accelerator ring, which
allows reliably separating the desired TVPC signal from the false contribution, as well
as from other signals using Fourier analysis of the measured counting rate of the number of events.
The same method can be used to measure the TVPC signal
in $^3$He$d$ scattering.

\textbf{3.
Elements of the ${N^3}$He and ${^3}$He${d}$ scattering formalism.}
Hadronic $T$-even $P$-even spin-dependent amplitudes of $pN$ scattering are chosen in the form
\cite{Platonova:2010wjt}
\begin{eqnarray}
M_N = A_N+C_N{\bfg \sigma}_p\cdot {\bf \hat n}+C'_N{\bfg \sigma}_N \cdot {\bf \hat n}\nonumber \\
 + B_N({\bfg \sigma}_p\cdot {\bf \hat  k})({\bfg \sigma}_N\cdot {\bf \hat  k})\nonumber \\
 +(G_N+H_N)(\bfg \sigma_p\cdot {\bf \hat q})(\bfg \sigma_N\cdot {\bf \hat q})\nonumber\\
 +(G_N-H_N)(\bfg \sigma_p\cdot {\bf \hat  n})(\bfg \sigma_N\cdot {\bf \hat n})
 \label{MpN};
 \end{eqnarray}
  here ${\bfg\sigma}_p$ (${\bfg\sigma}_N)$ are the Pauli spin matrices acting on the spin state of the proton beam (target nucleon $N$), the unit 
   vectors ${\bf\hat k}, {\bf\hat q}$ and ${\bf\hat n}$ are defined in terms of the initial ${\bf p}$ and final ${\bf p'}$ momenta  of the scattering proton:
 $\bf \hat k= ({\bf p+\bf p'})/{|\bf p+\bf p'|},\bf \hat q=  ({\bf p-\bf p'})/{|\bf p-\bf p'|},
  \bf \hat n=[ \bf \hat k\times  \bf \hat q]$.
In the Glauber theory, only $pN$ amplitudes on the mass shell contribute to the amplitude of $pA$ scattering.
We consider here the following three terms of the TVPC $t$-matrix  of elastic $pN$ scattering
that do not disappear on the mass shell
\cite{Beyer93,Uzikov:2015aua,Uzikov:2016lsc}:
 \begin{eqnarray}
t_{pN}=
 h_{p N}[({\bfg \sigma}_p\cdot {\bf \hat k})(\bfg \sigma_N\cdot {\bf \hat q})+
 ({\bfg \sigma}_p\cdot {\bf \hat q})({\bfg \sigma}_N\cdot {\bf \hat k})\nonumber \\
-\frac{2}{3}({\bfg \sigma}_N\cdot {\bfg \sigma}_p)({\bf \hat q} \cdot {\bf \hat  k})]
+g_{p N}[{\bfg \sigma}_p\times{\bfg\sigma}_N]\cdot [{\bf\hat  q}\times {\bf \hat k}] \nonumber \\
\times({\bfg \tau}_p-{\bfg \tau}_N)_z
+g_{p N}'({\bfg \sigma}_p-{\bfg \sigma}_N)\cdot i[{\bf \hat q}\times{\bf\hat  k}]
[{\bfg \tau}_p\times{\bfg \tau}_N]_z;
\label{tpN}
\end{eqnarray}
 here $h_{pN}$, $g_{pN}$, $g_{pN}'$  are the unknown amplitudes (constants) of TVPC $NN$ interactions,
${\bfg\tau}_p$ (${\bfg\tau}_N$) are the isospin Pauli matrices, acting on the state of the initial proton (nucleon $N$).

 The transition operator $N^3$He~$\to N^3$He ($N=p,n$) taking into account $T$-invariant
and $T$-non-invariant contributions has the same spin structure as the sum of the transition operators
$pN\to pN$ (\ref{MpN}) and (\ref{tpN}), since the spins of the initial and final particles in these processes are the same and equal to $s=1/2$:
\begin{eqnarray}
 F=A_1+A_2{\bfg \sigma}_N {\bf \hat n}+A_3{\bfg \sigma} {\bf \hat n}
 + A_4({\bfg \sigma}_N\cdot \bf \hat  k)({\bfg \sigma}\cdot {\bf \hat  k})\nonumber  \\
 +(A_5+A_6)(\bfg \sigma_N\cdot {\bf \hat q})(\bfg \sigma\cdot {\bf \hat q})\nonumber \\
 +(A_5-A_6)(\bfg \sigma_N\cdot {\bf \hat  n})(\bfg \sigma\cdot {\bf \hat n})\nonumber\\
 +h_{\tau N}[(\bfg \sigma_N\cdot {\bf \hat k})(\bfg \sigma\cdot {\bf \hat q})+
 ({\bfg \sigma}_N \cdot {\bf \hat q})({\bfg \sigma}\cdot {\bf \hat k})\nonumber\\
-\frac{2}{3}({\bfg \sigma}_N\cdot {\bfg \sigma})({\bf \hat q} \cdot {\bf \hat  k})]
+g_{\tau N}[\bfg \sigma_N\times{\bfg\sigma} ]\cdot [{\bf\hat  q}\times {\bf \hat k}]
\nonumber\\
+g_{\tau N}'i({\bfg \sigma}_N-{\bfg \sigma})\cdot [{\bf \hat q}\times{\bf\hat  k}]
[{\bfg \tau}_N\times {\bfg \tau}]_z;
\label{p3He}
\end{eqnarray}
here $\bfg\sigma$ ($\bfg\tau$) are spin (isospin) Pauli matrices
acting on the state of the $^3$He nucleus;
$A_i$ ($i=1,2,\dots,6$) are  $T$-even, and $h_{\tau N}$, $g_{\tau N}$, $g_{\tau N}'$ are
$T$-odd $P$-even amplitudes of the elastic $N^3$He scattering.
%
The
second-to-last term of the formula (\ref{p3He})
lacks an isospin factor similar to the factor in the formula (\ref{tpN}), which takes into account
 vanishing the $g$-type amplitude for scattering of identical nucleons. In the case of $N^3$He scattering, the colliding particles are not identical, and the amplitude of the $g$ type is nonzero for both the incident proton and the neutron.
Analytical expressions for all spin
amplitudes in (\ref{p3He}) are obtained by us in the framework of Glauber theory
using elementary $pN$ amplitudes (\ref{MpN}) and
(\ref{tpN}) for all three rescattering multiplicities, taking into account the spin structure of the $^3$He nucleus in the $S$-wave approximation for the spatial part of the wave function.

The spin structure of the 
transition operator for elastic $^3$He$d$ scattering is the same as for $pd$ scattering,
and for the case of collinear kinematics is given in Ref. \cite{Uzikov:2015aua}.
 The TVPC
effect in $^3$He$d$ scattering
is determined by the imaginary part of the TVPC amplitude of $^3$He$d$ scattering at zero angle
 $\tilde g$
  \cite{Uzikov:2015aua,Uzikov:2016lsc}:
 \begin{equation}
 \sigma_{TVPC}=-4\sqrt{\pi}\frac{2}{3}Im \tilde g.
 \end{equation}
  To calculate the $\tilde g$  amplitude, we use the Glauber theory.
In the single-scattering approximation, this amplitude vanishes due to the properties of the operators
  (\ref{tpN}) and (\ref{p3He}).
Given the compactness of the $^3$He nucleus,
we calculate this amplitude 
for the process $^3$He$d\to ^3$He$d$
similarly to that for
$pd \to pd$, based on the mechanism of
double scattering, with the difference 
 that instead of $pN$ amplitudes
the amplitudes of $^3$He$N$ scattering are included. Note that
in $^3$He$d$ scattering there are mechanisms of higher multiplicity,
but their contribution at zero scattering angle is expected to be significantly less than
that of the double-scattering mechanism, so, they are not considered in this paper.
Within the approximation of the double $^3$He$N$ scattering mechanism,
from Eq. (10) of Ref. \cite{Uzikov:2016lsc},
 we obtain the desired $\tilde g$ amplitude for the process $^3$He$d\to ^3$He$d$ in the
 following form:
 \begin{eqnarray}
  {\tilde g}= \frac{i}{4\pi m_p}\int_0^\infty dq q^2[ S_0^{(0)}(q)\nonumber \\
  -\sqrt{8}S_2^{(1)}-4S_0^{(2)}(q)
  +9S_1^{(2)}(q)
  +\sqrt{2}\frac{4}{3}S_2^{(2)}(q)]\nonumber \\
 \times \{-A_3^{\tau n}(q)h_{\tau p}(q)+A_3^{\tau p}(q)
  [g_{\tau n}(q)-h_{\tau n}(q)]\},
  \label{gtilde}
 \end{eqnarray}
where  $S_n^{(m)}(q)$ $(m,n=0,1,2)$ are the
deuteron form factors defined in
\cite{Uzikov:2016lsc},
taking into account the contribution of $S$ and $D$ waves, $u(r)$ and $w(r)$, respectively,
while the superscript $m$ indicates the degree of 
the $D$-wave contribution, $w^m(r)$,
and the 
subscript $n$ is the order of the spherical Bessel function $j_n(qr)$ 
under the integral over $r$.
\begin{figure}
    \centering
\vspace {0.5cm}
    \includegraphics[width=0.7\linewidth]{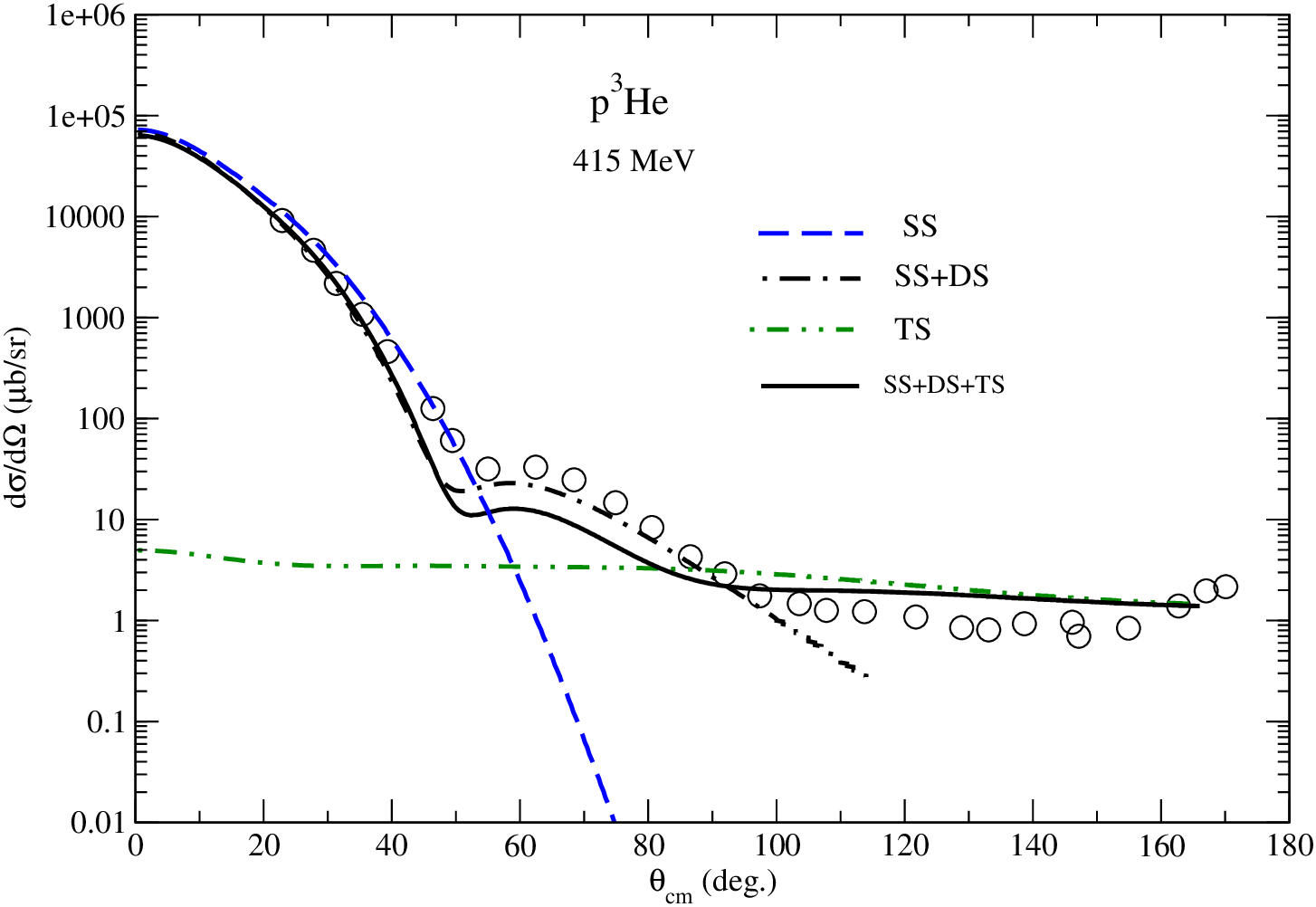}
    \caption{
Differential cross section of
$p^3$He elastic scattering at energy $T_p = 415$ MeV.
The contributions of single (dashed curve, SS)
and triple 
(dash-dot-dotted curve, TS) scattering are shown. The dash-dotted
curve (SS+DS) shows the summed contribution of single- and double-scattering, and the solid
curve shows the full calculation
(the sum of all three rescattering multiplicities).
Points  are  experimental data from the work
\cite{Hasell:1985gi}.}
    \label{fig:DS-515}
\end{figure}
\vspace{0.5cm}
\begin{figure}
    \centering
\vspace{0.5cm}
\includegraphics[width=0.7\linewidth]{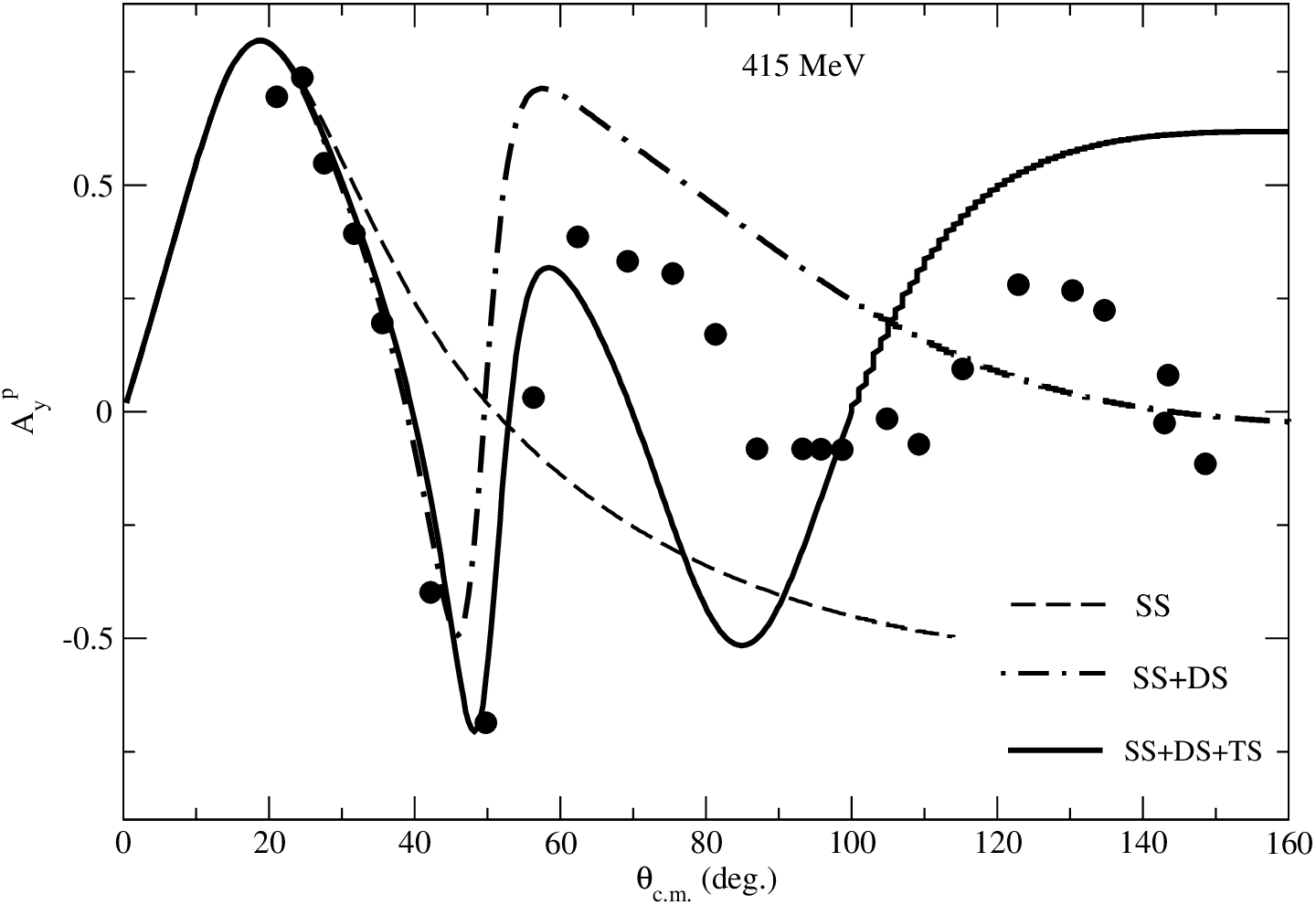}
    \caption{
 Proton vector analyzing power
in $p^3$He elastic scattering at energy $T_p = 415$ MeV. For a description of points and curves, see the caption of
 Fig. 1. }
    \label{fig:3}
\end{figure}

Let us consider separately the various $T$-odd contributions to the
TVPC signal in  the $^3$He$d$ scattering.
 i)
 As shown in \cite{Uzikov:2015aua,Uzikov:2017dns},
the contribution of the $g'$-term to the TVPC signal of $pd$ scattering
disappears due to the symmetry property of the charge-exchange transition
$\textlangle pn|\hat g'|np\textrangle=-\textlangle np|\hat g'|pn\textrangle$.
  The charge-exchange amplitude  $p^3$H~$\leftrightarrow n^3$He of the $g'$-type has a similar symmetry property, and therefore its contribution to the TVPC signal of the  $^3$He$d$ scattering also vanishes for all multiplicities of $pN$ scattering.
ii)
The contribution of the $g_N$-type interaction to the elastic $n^3$He scattering,
which is included in the TVPC signal (\ref{gtilde}),
vanishes for single- and double-scattering mechanisms, due to the specific spin properties of the transition operator
 with this interaction.
       In this case, the non-vanishing contribution is given by the amplitude of
       the triple scattering only, in which the $g_N$ terms enter in a superposition
with $h_N$ 
ones. Due to the influence of the $^3$He-nucleus form factor,
the contribution of triple scattering compared 
to that of single scattering for TVPC interaction is also suppressed, as for $T$-even $P$-even forces.
 iii) The $h_N$-type interaction makes a nonzero contribution to the amplitudes of
all three multiplicities of elastic $N^3$He scattering.

 Expressions for $p^3$He amplitudes $h^{(1)}_{p\tau}$ and $h^{(2)}_{p\tau}$ of the $h$ type in the approximation of
 single and double scattering, respectively,
are the following:
\begin{eqnarray}
 h^{(1)}_{p\tau}=\frac{k_{p\tau}}{k_{pN}}S(q)h_n(q), \nonumber \\
 h^{(2)}_{p\tau}=\frac{k_{p\tau}}{\pi ik_{pN}^2}S\left(q/2\right)
\int d^2q'S\left(\sqrt{3}q'\right)A_p(q_1)h_n(q_2),
 \label{h12}
\end{eqnarray}
where the vectors ${\bf q}_1$ and ${\bf q}_2$ are defined as
\begin{eqnarray*}
{\bf q}_1={\bf q}/2-{\bf q}', {\bf q}_2={\bf q}/2+{\bf q}',
\end{eqnarray*}
$S(q)$ is the  elastic $^3$He form factor: $S(q)=\exp{\{-q^2/12c^2\}}$,
and $k_{pN}$ ($k_{p\tau}$) is the relative momentum in the $pN$ ($p^3$He) system.
%
As the transferred momentum $q$ increases, both amplitudes decrease rapidly
 in magnitude due to
the $^3$He form factor,
therefore, the dominant contribution to the integral (\ref{gtilde}) is made by
the region of small transferred momenta and,
accordingly, by the single-scattering amplitude.
We calculate the amplitude of the triple scattering by taking the product  of the
$NN$-amplitudes  $f_1({\bf q}_1)f_2({\bf q}_2)f_3({\bf q}_3)$
out of the integral
over intermediate momenta at
${\bf q}_1={\bf q}_2={\bf q}_3={\bf q}/3$,
that leads to the following expression:
 \begin{eqnarray}
 h^{(3)}_{p\tau} &=& -\frac{1}{3}\frac{k_{p\tau}}{4\pi^2 k_{pN}^3}{\tilde S}
\{{ \Sigma}_p^2 h_n + (B_p+G_p+H_p) \nonumber \\
&& \times[(B_n+G_n+H_n)h_p + 2(B_p-G_p-H_p)g_n]\},
 \label{h3}
 \end{eqnarray}
where ${\tilde S}=\frac{64}{3}\pi^2c^4$,
\begin{eqnarray}
\Sigma_p^2=3A_p^2+C_p^2-3C_p'^2-2B_p^2-3G_p^2-3H_p^2-2G_pH_p.
\label{Sigm3}
\end{eqnarray}
 All $NN$-amplitudes in Eqs. (\ref{h3}) and (\ref{Sigm3})
are taken at the transferred  momentum value $q/3$.

 Note that in the amplitude of the triple scattering
(\ref{h3})
a $g_n$-type interaction contribution appears. This term is suppressed by the product
of spin-dependent amplitudes $B_p, G_p, H_p$ of the
$T$-even $pp$ scattering compared to the $h_n$ term multiplied by the spin-independent term $A_p^2$ in
Eq. (\ref{h3}). As will
be seen below from numerical calculations, the contribution of triple scattering of the $h$-type is significantly less than the contribution of single scattering.

The amplitudes of $n{}^3$He scattering, both $T$-even and $T$-odd, are obtained from the respective amplitudes of $p{}^3$He scattering by replacing $pp$ amplitudes in them by $np$ 
ones, and
$pn$ amplitudes by $nn$ ones.
So, the expressions for the $h$-type $n{}^3$He amplitude in the single- and double-scattering approximation
have the form similar to the expressions (\ref{h12}) with the replacement of indices $p \leftrightarrow n$.
To properly account for the $g$-type $NN$ amplitudes when calculating the $N{}^3$He amplitudes,
it is also necessary to take into account the isospin factor (see Eq. (\ref{tpN}))
vanishing for the collision of identical nucleons.
Accordingly, in the final formulas we assume $g_p=0$ and get that the $n{}^3$He amplitudes of $g$ type for single- and double-scattering mechanisms turn to zero, whereas the expressions for the $n{}^3$He amplitudes of $h$ and $g$ types in the triple scattering approximation have the form:
\begin{eqnarray}
\label{eq-hgn}
 h^{(3)}_{n\tau} &=& -\frac{1}{3}\frac{k_{n\tau}}{4\pi^2 k_{nN}^3}{\tilde S}
\{\Sigma_n^2 h_p + (B_p+G_p+H_p)\nonumber \\
&& \times[(B_n+G_n+H_n)h_n - 2(B_n-G_n-H_n)g_n]\}, \nonumber \\
 g^{(3)}_{n\tau} &=& -\frac{1}{3}\frac{k_{n\tau}}{4\pi^2 k_{nN}^3}{\tilde S}
\{ [B_n^2-(G_n+H_n)^2]h_p  \nonumber \\
&&+ (B_p-G_p-H_p)[(B_n+G_n+H_n)h_n \nonumber \\
&&- 2(B_n-G_n-H_n)g_n]\},
\end{eqnarray}
where $\Sigma_n^2$ is defined by the formula (\ref{Sigm3}) with the replacement $p\to n$.

\textbf{4. Numerical results and discussion.
}
 To test the model 
 developed for elastic $p^3$He scattering, we calculate the differential
cross section $d\sigma/d\Omega$ and the vector analyzing power  $A_y^p$ of this process:
\begin{eqnarray}
 \frac{d\sigma}{d\Omega} =
 \Sigma=|A_1|^2+|A_2|2+|A_3|^2+|A_4|^2 \nonumber \\
 +|A_5+A_6|^2+|A_5-A_6|^2,\\
 \label{difsec}
 A_y^p=2{\rm Re}[A_1A_2^*+(A_5-A_6)A_3^*]\Sigma^{-1}.
\label{Ayp}
\end{eqnarray}
%
Note that the amplitude $A_3$ included in the expression for the TVPC signal (\ref{gtilde})
enters with its phase the expression (\ref{Ayp}) for $A_y^p$, and 
its modulus squared enters the cross section.
%
%
In numerical calculations, we use the $pN$ scattering amplitudes from the SAID database~\cite{Arndt:2007qn}. For the $^3$He nucleus, we use a completely antisymmetric wave function with a symmetric
spatial $S$-wave function $\psi=N\exp{\{-c^2(r_1^2+r_2^2+r_3^2)\}}$, where the parameter
$c^{-1}=1.56\sqrt{2}$ fm \cite{Kondratyuk83}, and $N$ is a normalization factor.
%
 The deuteron wave function is taken in the CD-Bonn model of the $NN$ interaction potential
 \cite{Machleidt:2000ge}.


%
The results of calculations of the differential cross section and vector analyzing power $A_y^p$ for
$p^3$He elastic scattering at proton beam energy $T_p = 415$ MeV are shown in Figs. 1 and 2,
respectively, in comparison with the data from Ref. \cite{Hasell:1985gi}.
%
%
It can be seen from the figures that in the range of scattering angles from zero to $\sim 50^\circ$,
the sum of the amplitudes of single- and double-scattering mechanisms agrees well with experimental data.
The contribution of triple scattering becomes dominant in the cross section and $A_y^p$
at angles greater than $\sim 70-80^\circ$, 
where the cross section decreases by four orders of magnitude, and the area of these
large angles is located
beyond the scope of applicability of the Glauber theory.
We found that a similar situation takes the place
also at other proton energies -- 156, 200, 300, 515 MeV, where
the data are available \cite{Hasell:1985gi} for $d\sigma/d\Omega$ and $A_y^p$, as well as at energy of
1 GeV, for which only the differential cross section was measured. The corresponding formalism and
numerical results will be published 
elsewhere.

Obtained agreement with the data on $p^3$He scattering  in the forward  hemisphere
gives reason to believe that the calculation of the 
$T$-invariance violation
 effect in this approach
 is performed with a similar degree of accuracy 
(up to unknown TVPC constants).
  The results of calculation for  
the TVPC signal in
$^3$He$d$ scattering for $T$-odd interactions of $h$ and $g$ types with account of
mechanisms with different rescattering multiplicities are shown in Fig. 3 in dependence on the beam energy.
It can be seen from the figure that the mechanism of single scattering in
the $N^3$He$\to N^3$He  process completely dominates this signal for the $h_N$-type interaction. The $g_N$-type interaction gives
contribution only to the triple-scattering amplitude of the $n^3$He~$\to n^3$He process, which is included in the expression
(\ref{gtilde}) for the forward scattering $^3$He$d$ amplitude, and therefore it is suppressed compared to the $h_N$-type interaction by 3--4 orders of magnitude (assuming the $h$ and $g$ constants are equal).
 As already noted,
$NN$ interaction of the $g'$ type does not contribute to
the TVPC signal of the $^3$He$d$ scattering process. It follows from the calculations that in
the considered energy range, 0.1--1 GeV, the TVPC effect is a smooth function of energy.

  Fig. 4 shows the contributions of $S$- and $D$-waves of the deuteron
 to the TVPC signal.
It can be seen that the $S$-$D$ interference
 is very significant, although the contribution
of the $D$ wave itself is small.
The pure $D$-wave contribution is 
small compared to the interference term 
proportional to
$u(r)w(r)$,
due to the fact that the two form factors defined by the square of $D$ wave, $S_0^{(2)}$ and $S_2^{(2)}$,
 enter the TVPC signal with a minus sign, while the other form factor
$S_1^{(2)}$,  defined also by the square of $D$ wave, enters with a plus sign
(see \cite{Uzikov:2016lsc}), whereas the interference term, $S_2^{(1)}$, is
not a subject to such compensation.

\begin{figure}
    \centering
\vspace{0.5cm}
      \includegraphics[width=0.7\linewidth]{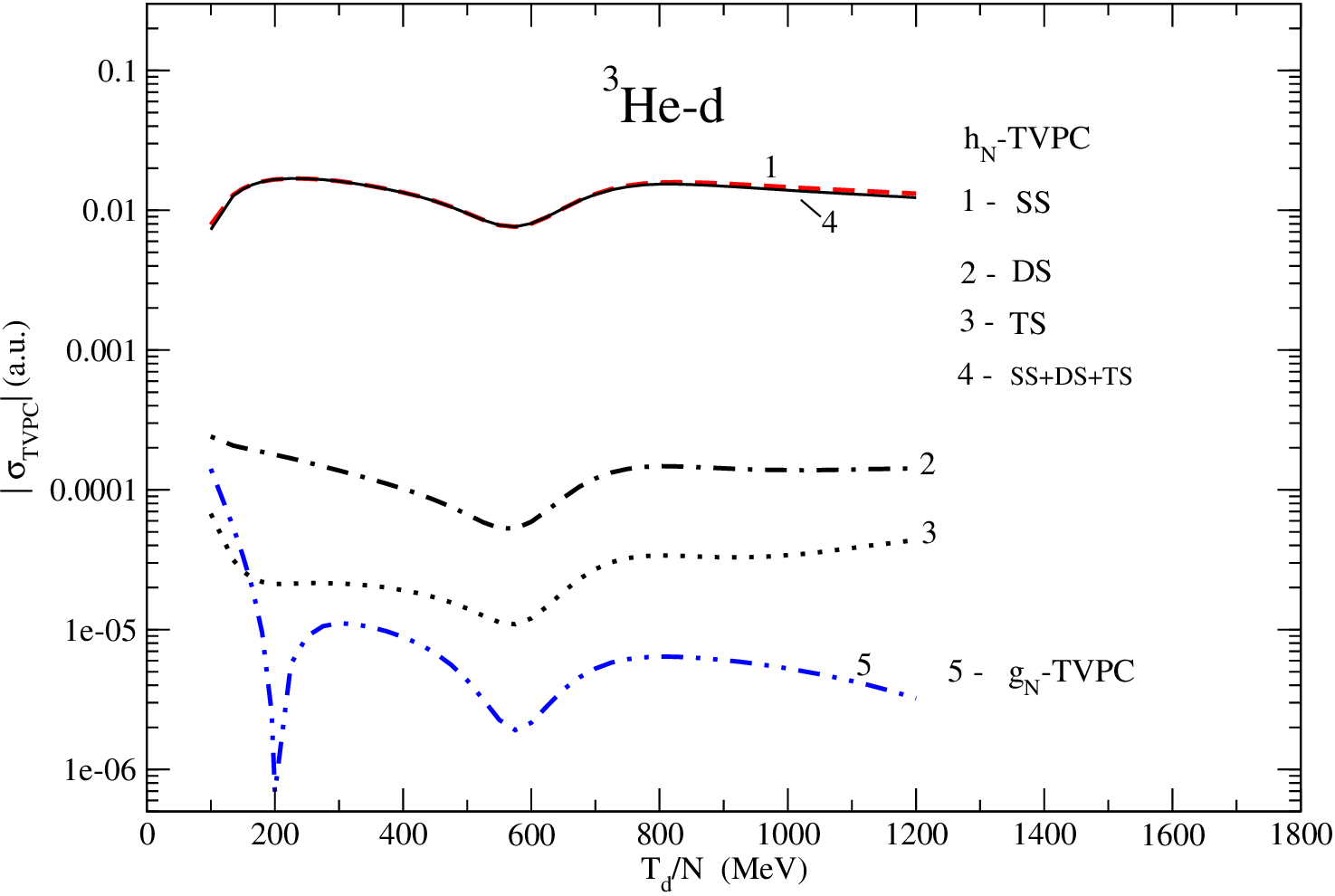}
    \caption{
 The TVPC signal in $^3$He$d$ scattering depending on the 
collision energy per nucleon
for $NN$ interactions of $h_N$ (1--4) and $g_N$ (5) types,
when the TVPC $NN$ amplitudes are included in single- (SS),
double- (DS) and triple- (TS)
scattering amplitudes of the $^3$He$N \to ^3$He$N$ process. 
The total calculation result (SS+DS+TS) is shown by the
curve 4 for $h_N$ and 5 for $g_N$ interactions.
    }
    \label{TVPC-hg}
\end{figure}

\begin{figure}
    \centering
 \vspace{0.5cm}
    \includegraphics[width=0.7\linewidth]{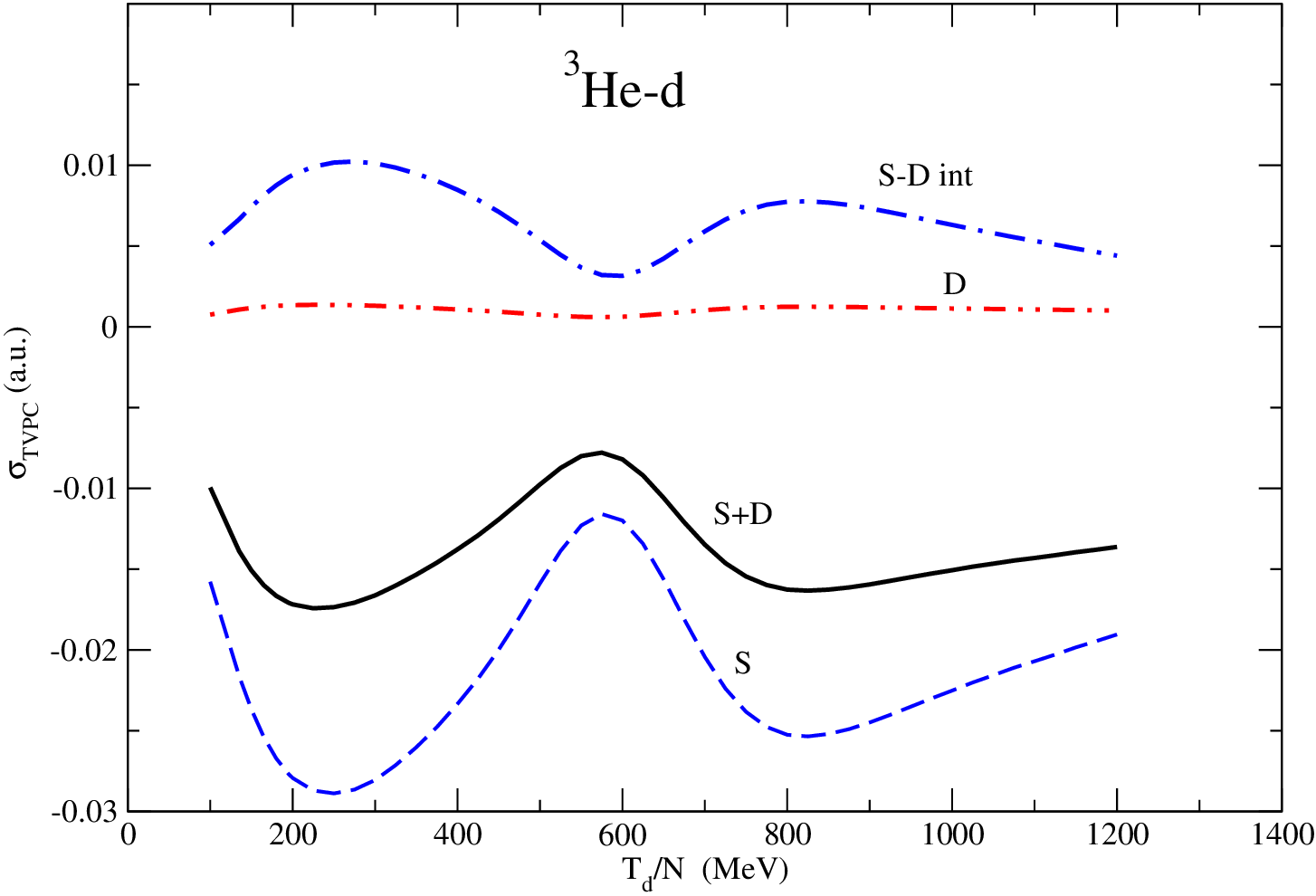}
    \caption{
The  TVPC signal of $h_N$ type
in $^3$He$d$ scattering
depending on the collision energy per nucleon, taking into account various deuteron
form factors $S_n^{(m)}$
in the expression (\ref{gtilde}) with the deuteron  $S$ and $D$ waves:
S -- $S_0^{(0)}$, D -- $S_0^{(2)}$, $S_1^{(2)}$ and $S_2^{(2)}$, S-D int -- $S_2^{(1)}$,
S+D is the total 
calculation result.
    }
    \label{fig:TVPC-SD}
\end{figure}

\textbf{5. Conclusion.}
The effect of
$T$-invariance violation while $P$-parity conserving  (the TVPC signal) in the process of scattering of polarized $^3$He nuclei
on tensor-polarized deuterons has been calculated on the basis of the Glauber multiple-scattering theory, taking into account the full spin dependence of $NN$ scattering amplitudes and including all $T$-odd terms non-vanishing on the mass shell.
The $T$-even as well as $T$-odd
spin-dependent amplitudes of elastic scattering of protons and neutrons on $^3$He
nuclei necessary for this calculation are constructed within the framework of the Glauber theory, taking into account the contributions of all three rescattering multiplicities in the $S$-wave approximation for the $^3$He nucleus wave function.
 Numerical calculations demonstrate quite good agreement with the available data on the differential cross section
and vector analyzing power  $A_y^p$ of $p^3$He elastic scattering at incident proton energies of 0.1--1 GeV in the forward hemisphere, at those scattering angles where the contribution of single-scattering  mechanism with an admixture of its interference
with double scattering dominates.

%
The calculation of the TVPC signal in $^3$He$d$ scattering shows that  the mechanism of single scattering in
the $N^3$He$\to N^3$He process   completely dominates  this signal for the
$h_N$-type interaction.
The $g_N$-type interaction gives
contribution only to the triple-scattering amplitude of the process $n^3$He$\to n^3$He
and therefore, it is significantly suppressed compared to the $h_N$ interaction
(with the $h$ and $g$ constants assumed to be equal).
%
$NN$-interaction of the $g'$ type does not contribute
to the TVPC signal of the process under consideration.

%
The dominant contribution of only one type of $T$-odd nucleon-nucleon forces found in
the calculated TVPC signal of the $^3$He$d$ scattering
process significantly simplifies the task
of extracting the unknown constant of the $h_N$ interaction from the corresponding data.
%
It should be noted that
experiments on the transmission of polarized neutrons through tensor-polarized nuclei are planned
at low energies under conditions when a significant increase of the TVPC effect is expected due
to the resonant properties of the nuclear structure~\cite{Bowman:2014fca}.
However, due to the exceptional complexity of the resonant structure of multi-nucleon nuclei,
it will be very problematic to determine the absolute value of the constants of the TVPC
interaction from the expected data,
in contrast to the few-nucleon systems considered here.

\textbf{Acknowledgements.}
This  study was carried out at the expense of the grant of the Russian Science Foundation No. 23-22-00123,
https://rscf.ru/project/23-22-00123 /.

\bibliographystyle{ieeetr}
\bibliography{uz-pl.bib}
\end{document}